%% file: main.tex
\documentclass{aastex63}
\usepackage{graphicx} % Required for inserting images
\usepackage{xcolor}
\begin{document}

%\title{\textcolor{blue}{RomAndromeda:} The Roman Survey of the Andromeda Halo}
\suppressAffiliations

 %       \normalsize
 %       \textbf{Submitting Authors}: 
 %    \input{subauth.tex}
%Arjun Dey $<$arjun.dey@noirlab.edu$>$,
%Joan Najita $<$ joan.najita@noirlab.edu $>$, 
%Carrie Fillion $<$ cfilion@jhu.edu $>$
%Jesse Han $<$jesse.han@cfa.harvard.edu$>$
%       \vspace{0.1cm}
        
%        \textbf{Affiliation}: NOIRLab, NOIRLab, Johns Hopkins University, Harvard-Smithsonian Center for Astrophysics
        
%        \textbf{Contributing Authors}
\input{subauth.tex}
%\input{auth.tex}
%\vspace{-0.8cm}

\begin{abstract}
    As our nearest large neighbor, the Andromeda Galaxy provides a unique laboratory for investigating galaxy formation and the distribution and substructure properties of dark matter in a  Milky Way-like galaxy. Here, we propose an initial 2-epoch ($\Delta t\approx 5$yr), 2-band Roman survey of the entire halo of Andromeda, covering 500 square degrees, which will detect nearly every red giant star in the halo (10$\sigma$ detection in F146, F062 of 26.5 26.1~AB mag respectively)
    %as well as 
    and yield proper motions to $\sim$25 microarcsec/year (i.e., $\sim$90~km/s) for all stars brighter than F146 $\approx 23.6$ AB mag (i.e., reaching the red clump stars in the Andromeda halo). This survey will yield (through averaging) high-fidelity proper motions for all satellites and compact substructures in the Andromeda halo and will enable statistical searches for clusters in chemo-dynamical space. Adding a third epoch during the extended mission will improve these proper motions by $\sim t^{-1.5}$, to $\approx 11$~km/s, but this requires obtaining the first epoch in Year 1 of Roman operations.  In combination with ongoing and imminent spectroscopic campaigns with ground-based telescopes, this Roman survey has the potential to yield full 3-d space motions of $>$100,000 stars in the Andromeda halo, including (by combining individual measurements) robust space motions of its entire globular cluster and most of its dwarf galaxy satellite populations. It will also identify high-velocity stars in Andromeda, providing unique information on the processes that create this population. %This dataset will be foundational in understanding the immigration history of Andromeda, the formation of its halo, and the underlying scaffolding of its dark matter. 
These data offer a unique opportunity to study the immigration history, halo formation, and underlying dark matter scaffolding of a galaxy other than our own.
\end{abstract}

%\eject

\begin{center}
        \large
       \textbf{RomAndromeda: The Roman Survey of the Andromeda Halo}

       \vspace{0.5cm}
       \normalsize
       \begin{flushleft}
        \textbf{Roman Core Community Survey Category:} High Latitude Wide Area Survey
        \vspace{0.1cm}
        
        \normalsize
        \textbf{Scientific Categories:} \textit{stellar physics and stellar types; stellar populations and the interstellar medium; galaxies; the intergalactic medium and the circumgalactic medium}
        \vspace{0.1cm}

        \normalsize
        \textbf{Submitting Authors}: Arjun Dey, Joan Najita, Carrie Filion 
        \vspace{0.1cm}

        \textbf{Emails}: arjun.dey@noirlab.edu, joan.najita@noirlab.edu, cfilion@jhu.edu
        \vspace{0.1cm}
    
        \textbf{Affiliation}: NOIRLab, NOIRLab, JHU
        \end{flushleft}

        \textbf{Contributing Authors}
   \end{center}
%\vspace{0.1cm}
\input{auth.tex}

\section{Science Case}
% \textcolor{blue}{\textbf{Text from Carrie in blue, hoping to avoid accidentally overwriting your work! VERY rough draft}}

Dark matter drives nearly every aspect of galaxy formation and evolution, yet its nature remains one of the most fundamental and persistent questions in astrophysics. In the $\Lambda$CDM paradigm, galaxies live in dark matter halos that form hierarchically, and their present-day state is determined by their merger histories. While dark matter is not directly observed, its influence on baryonic matter can be and observations of resolved stars provide a unique handle with which to set constraints on the properties of dark matter. For example, the motions of halo stars can reveal the spatial distribution of dark matter and the existence of dark-matter substructure. %Star-by-star kinematic (and chemo-dynamic) analyses have historically only been possible in the Milky Way and nearby dwarf galaxies, where there are large numbers of stars with proper motions and line-of-sight velocities. 
So far, such studies have only been feasible in the Milky Way and a handful of its dwarf galaxy satellites. However, with new ground-based spectroscopic facilities and the Roman telescope, it is now possible to perform such analyses in the Andromeda Galaxy (M31), the Milky Way’s nearest large %spiral galaxy 
galactic neighbor.

 %Located $\approx780$~kpc from the Milky Way, 
 Andromeda %has the potential to be
 is an ideal dark matter laboratory. Unlike the Milky Way, it offers an external view of a galaxy --
 %we have an external view of Andromeda and
 its  entire stellar halo is 
 spread out over only $\approx$~500 sq.deg.  %Its proximity 
Andromeda's proximity 
 ($\approx 780$~kpc away) allows individual stars to be resolved with sufficiently deep imaging and spectroscopy. As a result, extensive ground-based imaging and spectroscopy and space-based imaging studies of Andromeda have been undertaken. These hint at a complex formation history, in some ways similar to that of the Milky Way \citep[e.g., PAndAS, PHAT, SPLASH, etc.][]{Ferguson2002,Ibata2004, Brown2006, Koch2008, Guhathakurta2006, Kalirai2006a, McConnachie2018, Gilbert2009b, PHAT2012, PHAT2014,  Gilbert2019,  Escala2022, Dey2023}. Spectroscopic campaigns on new massively multiplexed spectrographs (e.g., Mayall/DESI; \citealt{Dey2023} and Subaru/PFS; \citealt{PFS_Science2014}), will soon 
 %result in 
 deliver line-of-sight velocity measurements for potentially $>10^5$ unique stars, but lack comparable tangential velocity information. 

The Roman Observatory provides the {\underline{\it only}} near-term opportunity  to map the entire Andromeda halo at $\approx$~0.1~arcsec resolution, and to measure proper motions for individual stars and substructures in the stellar halo. These data are critical to understanding galaxy formation. With Roman’s unique combination of wide-field, high spatial resolution, and low background, 
%enable the observation of 
it is now possible to observe
the entire Andromeda halo %including the identification of 
and identify
nearly every compact or tidally disrupted satellite within the system
and  discover new stellar streams in the halo.  Thin,  dynamically cold (low velocity-dispersion) stellar streams, such as those from tidally disrupted globular clusters, are of particular interest for constraining 
the nature of dark matter. 
% below added by SP
Roman can photometrically detect Palomar 5-like globular cluster streams in Andromeda \citep{pearson_2019,pearson_2022}, and (from the stream morphologies) can also detect gaps \citep{Aganze2023} in such streams induced by interactions with dark matter subhalos \citep[see e.g.,][]{yoon2011}. 
The morphology of streams and their underdensitites can help constrain the overall dark matter halo potential of Andromeda \citep{Nibauer2023}. In Andromeda we will have the advantage of searching for globular streams far from the galactic center, where streams are less susceptible to dynamical torques from baryonic perturbers associated with Andromeda's disk or bulge \citep[e.g.,][]{amorisco2016,pearson2017,banik2019}. 
The distribution of the properties of stellar substructures in the halo can be used to test the $\Lambda$CDM paradigm of structure formation \citep{Bell2008, Shipp2023}, as well as deconstruct the galaxy’s accretion history in terms of the characteristic epoch of accretion and the mass and orbits of progenitor objects \citep{Johnston2008, Sharma2011}.
Characterizing the population of dark matter subhalos with stellar streams allows constraints on the particle nature of dark matter \citep[e.g.,][]{Benito2020} in addition to constraining the overall halo structure.

%Morphology alone can provide valuable information: gaps, fans, or other anomalies along stream tracks indicate a history of interaction with dark matter substructure \textcolor{blue}{\textbf{not necessarily!! Resonances also cause such features}}. \cite{pearson_2019}  forecasts the ability of Roman to detect such streams. %, and HST has already demonstrated the feasibility of measuring motions for average populations at the distance of Andromeda \citep[e.g.,][]{Sohn2012}.

In addition, by obtaining two epochs of observations spread out over the 5-year baseline mission, Roman can provide proper motions for almost every spectroscopically observable Andromeda halo giant star, thus 
%allowing for 
enabling the first detailed 3-d dynamical study of the halo of a galaxy beyond the Milky Way. The survey, combined with ground-based spectroscopy, will yield the orbits of all the globular clusters and most of the Andromeda satellite galaxies, providing additional constraints on Andromeda's accretion history \citep[e.g.,][]{Mackey2019,Mackey2019b} and the \lq planes of satellites' problem (e.g., \citealt{ibata_2013}). These proper motions will be 
anchored by background QSOs and galaxies identified by the previous, current, and ongoing spectroscopy campaigns (as in \citealt{Sohn2012,Sohn2020}). 
For example, DESI is spectroscopically targeting more than 300 QSO candidates per square degree, and each Roman field should include $\sim$80 confirmed QSOs (i.e., stationary point sources) in addition to $\sim10^5$ faint galaxies to tie down the reference frame and measure differential proper motions for the stars, cf.~measurements using HST by \citet{Sohn2012}. %The baselines for the proper motions 
% Worried that this undermines the case for Roman: 
%The proper motion baselines can potentially be extended further by incorporating archival imaging. 
The wide-field imaging of the stellar halo from CFHT/PAndAS also provides a 
%potential 
$\approx$~20 year baseline for proper motions when combined with the proposed Roman data. Similarly, in Andromeda's disk, the combination of Roman and existing HST/PHAT survey data can provide more accurate stellar motions in the disk of the galaxy, as well as 
%potentially 
a 3-d reddening map of the Andromeda disk. These proper motions, combined with spectroscopy (from e.g. JWST or DESI), can enable analyses of the chemo-dynamical trends of different populations in the disk, probing its evolution.

\begin{figure*}[t]
    \centering
    \includegraphics[width=0.45\textwidth]{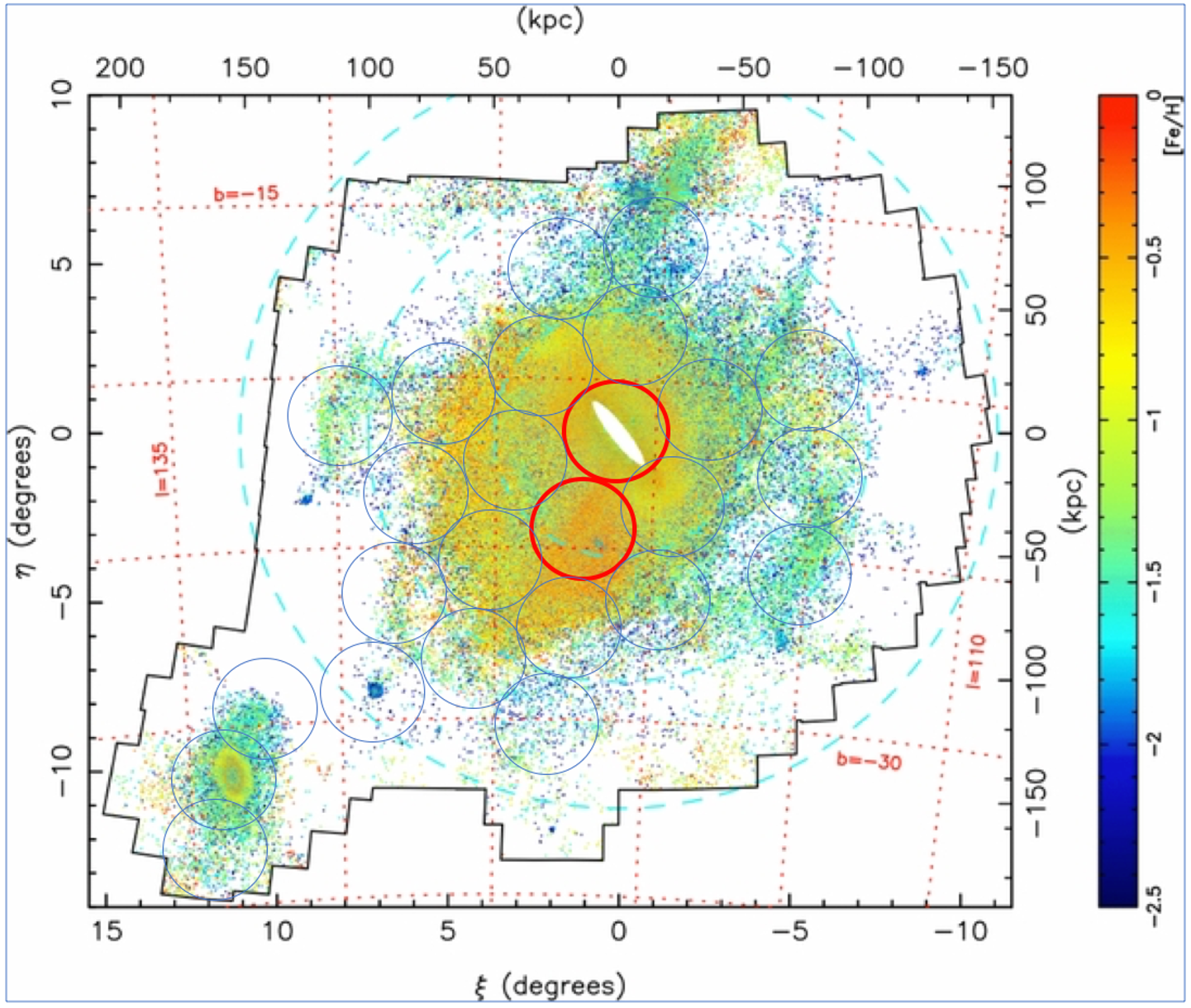}
    \caption{The metallicity distribution of stars in the Andromeda halo and the Triangulum galaxy \citep[from][]{Ibata2014} estimated using photometric data from the PAndAS survey \citep{McConnachie2018}. The proposed Roman survey will encompass the entire 500 square degree area shown in this figure, reaching beyond the outline of the PanDAS field (solid black line). The dashed cyan circles are at projected radii of 50, 100, and 150~kpc. Much of the halo can be covered by spectroscopic observations using the Mayall Dark Energy Spectroscopic Instrument (DESI; potential pointings shown in the blue circles) and the Subaru PFS surveys. Image credit: \citet{Ibata2014}}
    \label{fig:pandasmap}
\end{figure*}

\begin{figure*}[t]
    \centering
    \includegraphics[width=0.45\textwidth]{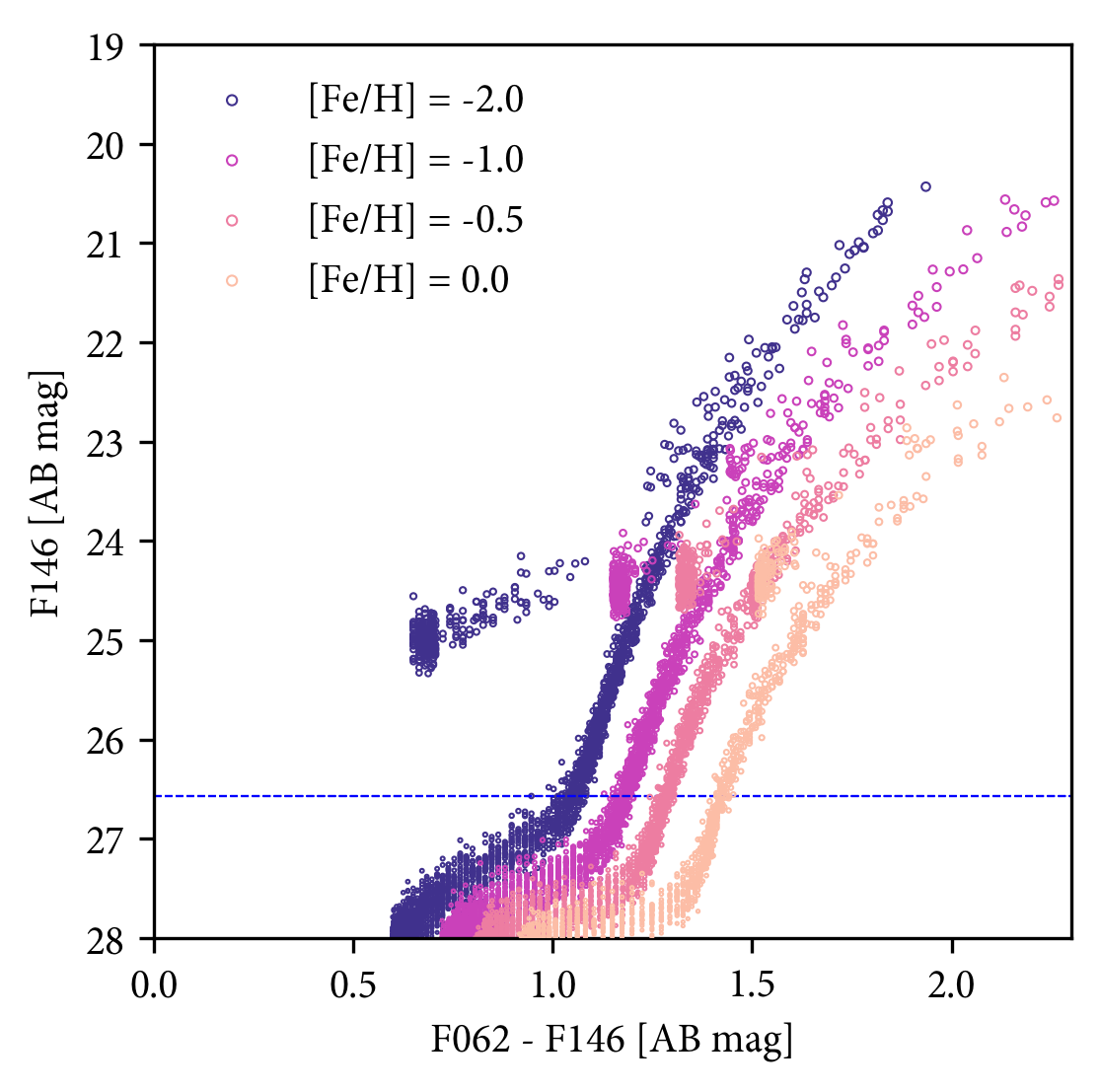}
    \includegraphics[width=0.45\textwidth]{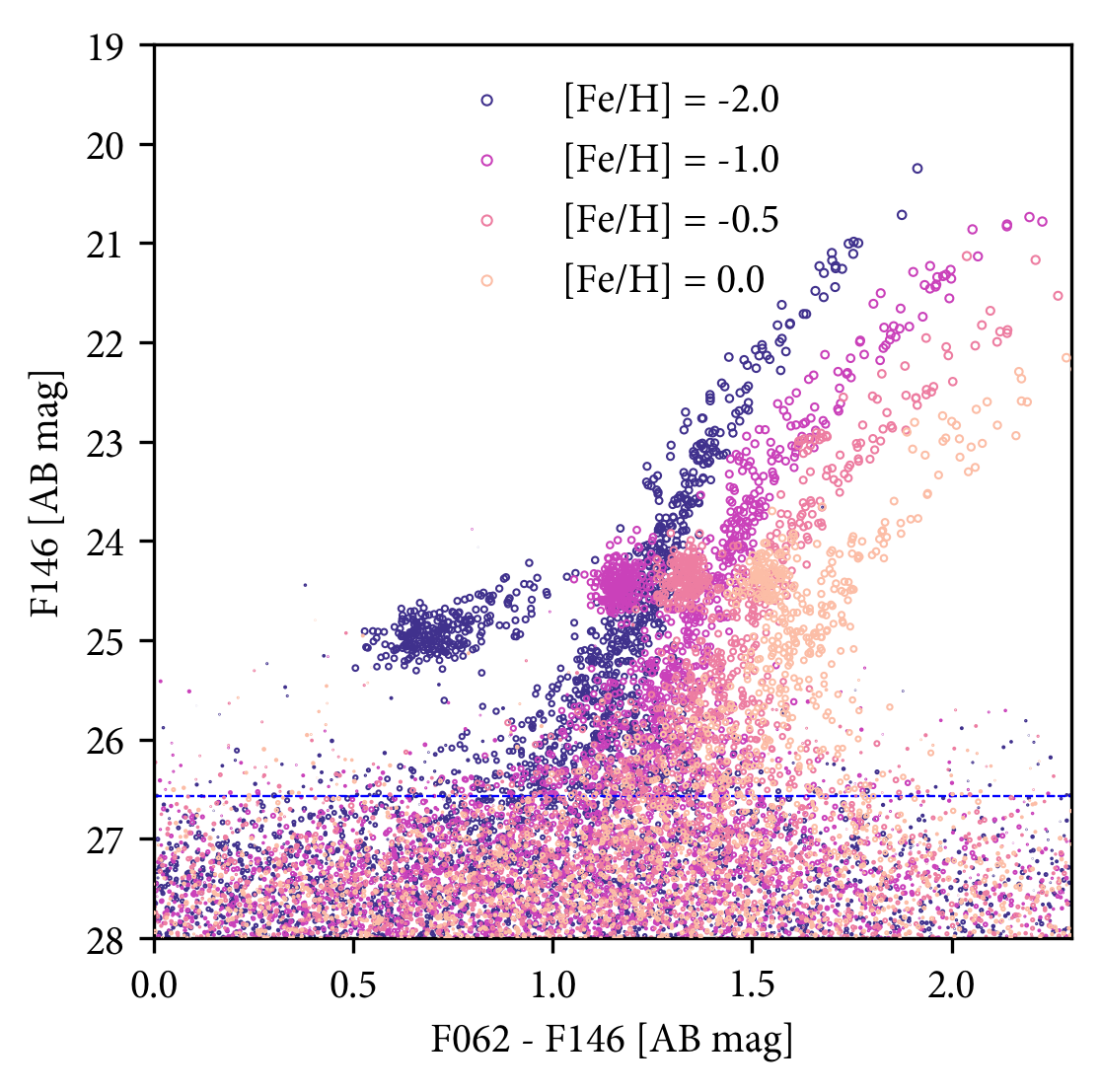}
    \caption{{\it Left:} Color-magnitude diagram for stars with a range of metallicities distributed at distances of 780$\pm$50~kpc in Andromeda's halo. The horizontal dashed line represents the 10$\sigma$ depth of the proposed survey achieved for the full survey (14x30 sec exposure time). {\it Right:} The photometric color-magnitude diagram (with expected scatter) resulting from the proposed Roman observations will be sensitive to nearly the entire red giant branch across the entire Andromeda halo. Thus, it will be capable of identifying nearly all the faint satellite companions (including ultra-faint dwarfs) and stars from most of the tidally stripped streams originating from clusters $M_V\approx-5$ and brighter. The photometry are computed using the MIST isochrones  \citep{MIST2016} and converted to AB mags assuming offsets of 0.122 and 0.993 mag for F062 and F146 respectively (See \url{http://mips.as.arizona.edu/~cnaw/sun.html}; \citealt{Willmer_2018}). Figure credit: Jesse Han.}
    \label{fig:cmd}
\end{figure*}

\section{Survey Strategy}

We propose a 500 square degree Roman Wide Field Instrument (WFI) public survey of the region covering the entire halo extent of the Andromeda and Triangulum galaxies (see Figure 1) in the F062 and F146 filters. 
WFI can cover this area in a single non-overlapping tiling of 1785 fields. The F146 filter provides the most efficient option for deep imaging due to its width. F062 provides the longest possible color baseline from F146, and therefore the best option for measuring the optical-to-near-infrared colors for the individual stars. In addition, F062 provides the smallest PSF, which when sampled with the multiple passes, can help us improve the centroids derived from the F146 data. 

We propose a total of 21 passes over this field, 14 dithered observations in the F146 filter (split into two epochs of 7 dithers each), and 7 dithered observations in the F062 filter. For an individual exposure time of 30 sec per observation, each tiling would require 1785 x (30 sec exposure + 50 sec slew) $\approx$~40 hours/tiling; thus, the entire survey would require a total of $\approx$~35 days of observation. Each epoch of $7\times30$ sec will result in a 10$\sigma$ detection of $\approx$~26.1 AB mag in both the F062 and F146 bands (see Table 1). This will reach every red giant star in Andromeda with absolute magnitudes $M_{({\rm F146,F062})} \le +1.5$ AB mag. This depth is nearly 1.5-2 mag deeper in F146 than a typical  Andromeda red clump star (see figure 2), and should enable detection of every ultra-faint dwarf galaxy to $-4$ to $-4.7$ mag (for detecting 10 or 20 stars per UFD). Even in Andromeda's bright disk, $\sim$20 arcmin from the center, Roman's excellent spatial resolution will allow us to reach the horizontal branch before the photometry is limited by crowding, as determined by applying the method of \citet{Olsen2003} to the surface brightness data from \citet{Jarrett2003}. In the bulge, $\sim$2 arcmin from the center, we will resolve the brightest individual red giants. To put this in context, the proposed Roman survey of Andromeda will provide a depth equivalent to what was achieved by the Sloan Digital Sky Survey at a distance of 100~kpc in the Milky Way Galaxy! 

Since deriving constraints on the proper motions of Andromeda stars, clusters, and satellite galaxies are the most critical part of the science, we propose to maximize the time baseline by scheduling the two epochs in years 1 and 5 of Roman science operations. In the event that there is an extended mission, we propose adding a third epoch later to extend the time baseline. Color information is critical to identifying clusters and coherent streams and determining their distances, and hence we propose to obtain imaging in 2 filters during the first epoch. This implies a time commitment of 23.14 days in year 1 and 11.57 days in year 5.

\begin{table}[t]
\begin{center}
\begin{tabular}{lccccc}
\hline
Mode &  Exposure Time & \multicolumn{2}{c}{F146} & \multicolumn{2}{c}{F062} \\
& (sec)  &  5$\sigma$ & 10$\sigma$ &  5$\sigma$ & 10$\sigma$ \\
\hline
\hline
Single Exposure & 30 & 25.32 & 24.52 & 24.82 & 23.92 \\
Single Epoch & 210 & 26.92 & 26.13 & 26.95 & 26.10 \\
Combined 2-epoch & 420 & 27.35 & 26.57 & -- & -- \\
\hline
\end{tabular}
\end{center}
\caption{Point-source depths (in AB mags) reached by the proposed Roman survey of Andromeda for single exposures, the combined first epoch observations (i.e., 7 $\times$ the single exposure depth), and the full 2-epoch depth. Including the slew time overheads, the entire proposed survey takes a total time of $\approx$~35 days. Depths were computed using the Roman ETC Jupyter Notebook and assuming typical Zodiacal light contributions 1.4$\times$ the average.}
\label{table}
\end{table}

The number of passes is driven by both astrometric and depth requirements. The Roman PSF is undersampled at the 0.11 arcsec pixel scale, and an accurate reconstruction of the PSF for centroiding at each epoch requires a large number of dithers. Rather than small dithers at each position, we propose 14 separate (but spatially staggered) tilings for the first epoch (7 in F146 and 7 in F062) in year 1, and then the last 7 tilings (in F146) during the second epoch in year 5. This has the advantage of providing a photometric time sequence (7 separate photometric points in each filter) during the first astrometric epoch which would allow for the identification of variable stars, especially RR Lyrae and Cepheids, for distance measurements and anchoring the distance scale for cosmological studies.  The availability of color information in the first year will allow the community to work on identifying all the substructure in the Andromeda halo, and thus mount more targeted spectroscopic campaigns for measuring line of sight  velocities and abundances. The total of 14 (7) passes in the F146 (F062) filter should allow for super-sampling of the PSF and thus more accurate per-epoch positions. The combination of the F062 and F146 data (and the existing ground-based imaging and imminent spectroscopic data) will allow us to model the spectral energy distribution through the wide F146 band for more accurate extraction of colors. 

\section{Deliverables}

The main data deliverables of the survey are: (1) photometry in two bands for all stars and galaxies down to $\approx$~26 AB mag (10$\sigma$); (2) proper-motion measurements or constraints for all stars brighter than $\approx$~23.5 mag (due to the higher signal-to-noise constraints for proper motion measurements); (3) per-epoch photometry for all sources detected at $>5\sigma$ (i.e., with F146, F062 $< 25.3, 24.8$ AB mag) in the individual epoch data. The photometric measurements and short-timescale single-epoch measurements will be available immediately after the first year. The proper motion catalog and long-term variability data will be available after the last epoch. 

With this plan, significant progress can be made immediately after the first year data are taken: e.g., measurements of the spatial distribution of halo stars; the photometric metallicity distributions; determination of the halo extent; source classification and star-galaxy separation; identifications of all overdense stellar substructures, i.e., clusters, dwarf galaxies and stream candidates; identification of nearly all short-term variables (e.g., RR Lyra, Cepheids, etc.). The addition of the second epoch enables the determination of proper motions for large numbers of stars in the halo and disk of Andromeda, and promises to revolutionize the study of Andromeda much in the same way that the \textit{Gaia} astrometric data revolutionised dynamical analyses of the Milky Way. 

%\textbf{Proper motions will also help with stream detection (e.g. streamfinder from Malhan and Ibata 2018) + verification, can help constrain orbits of satellite galaxies}

% \textcolor{blue}{\textbf{First Epoch Science:} With the first epoch of imaging, it will be possible map the density structure of the halo and disk of Andromeda. These data will enable the identification of over-dense stellar substructures, such as streams and shells, down to lower surface-brightness limits than previously possible. Indeed, the success of the PAndAS and PHAT imaging campaigns reveal the power of photometric analyses of Andromeda, and PAndAS exemplifies the science that is possible with photometry in just two filters. Analyses of the PAndAS data have revealed the existence of a number of stellar substructures \textbf{cite PAndAS substructure papers}, and by pushing down the stellar luminosity function, Roman promises to extend such analyses to even fainter substructures.}

%{\textbf{Second Epoch Science:} The addition of the second epoch enables the determination of proper motions for large numbers of stars in the halo and disk of Andromeda. The addition of proper motions promises to revolutionise the study of Andromeda much in the same way that the \textit{Gaia} astrometric data revolutionised dynamical analyses of the Milky Way...
%These star-count based analyses can identify breaks in the halo density, similar to analyses performed in the Milky Way \textbf{cite Jesses paper, Deason papers, Xue paper, etc...}

\begin{figure*}[t]
    \centering
    \includegraphics[width=0.85\textwidth]{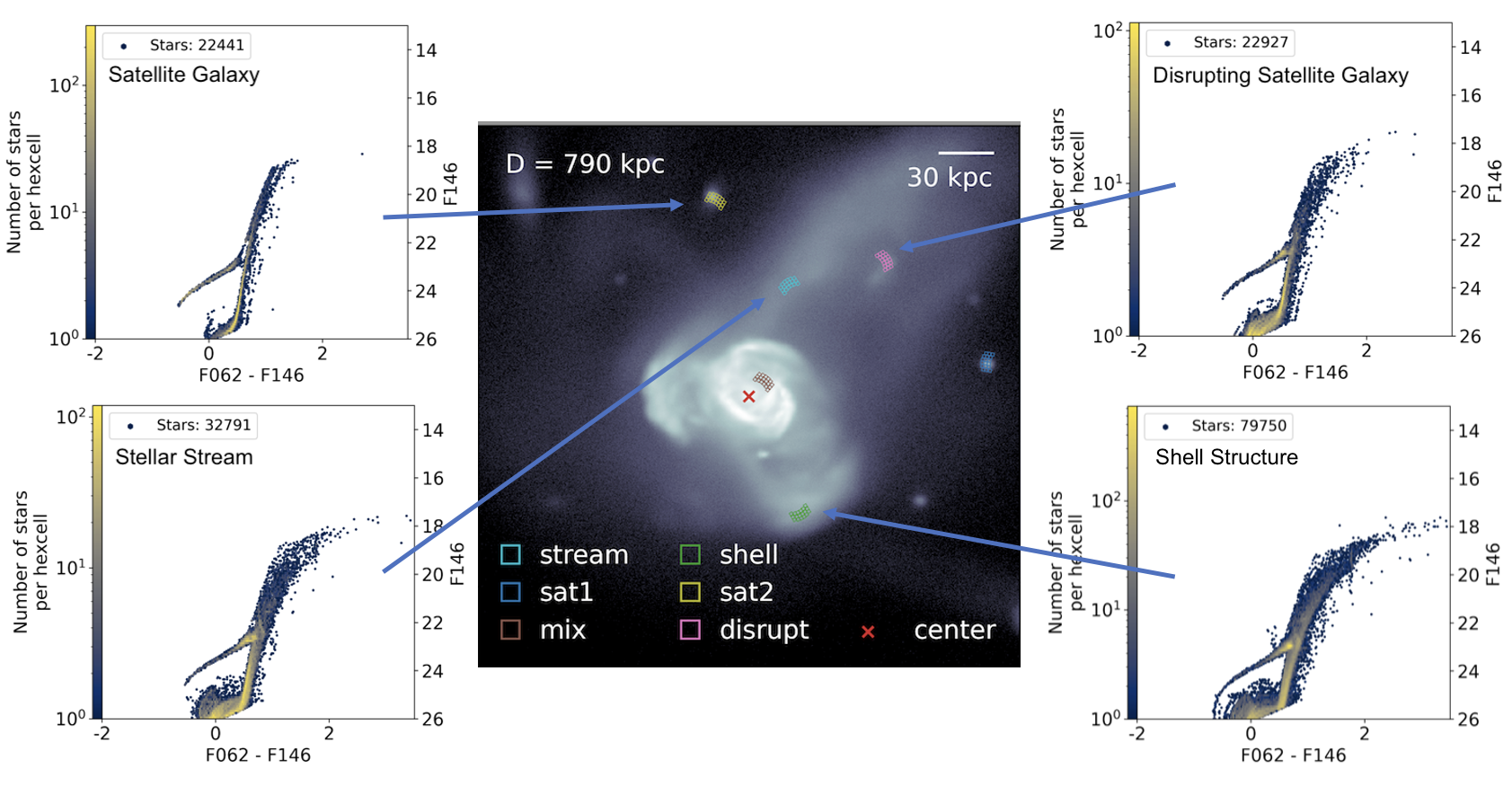}
\caption{Mock observations with Roman of a simulation of a galaxy with a mass similar to that of Andromeda and observed from a comparable distance. The central image shows the stellar number density for a mock stellar catalog, limited to a 10$\sigma$ detection in F146, F062 of 26.5, 26.1~AB mag, in which substructure is evident even by-eye. The example color-magnitude diagrams shown in the four panels surrounding the density image are constructed from the regions covered by single Roman WFI pointings and represent $\approx$~0.22\% of the data from the proposed Roman survey. The model galaxy is from the FIRE simulation suite \cite[{\tt m12f},][]{Garrison-Kimmel2017} and the stellar catalogs were generated using the {\tt py-ananke} software (\url{https://github.com/athob/py-ananke}) with isochrones for Roman photometry queried from the Padova/CMD database (\url{http://stev.oapd.inaf.it/cmd}). Figure credit: Adrien Thob.}
    \label{fig:simulation}
\end{figure*}

\section{Possible Implementation Options}

\begin{itemize}
    \item[\bf{(I)}] Cover Andromeda as part of the High Latitude Survey
    \item[\bf{(II)}] Design a custom survey of Andromeda as part of the Core Community Survey plan
\end{itemize}

\noindent{\bf Option I:} The notional plan for the High-Latitude Wide-Area Survey (HLWAS) will cover $\approx$~2000 square degrees in 4 bands (F106, F129, F158, F184) to depths of 25.8 to 26.7 AB mag. While this plan is much more extensive than that proposed here for Andromeda, at least half of the Andromeda halo footprint proposed here (e.g., the fraction which avoids the disks of Andromeda and Triangulum and lies at Galactic latitude $< -20$) could be completed as part of the HLWAS and attain the same goals. The pros of this implementation strategy would be: (a) larger number of (narrower) filters, which result in % better 
additional colors for stars and less PSF variation within each filter; (b) many more epochs than proposed here, which will improve the detection of variable stars and allow us to solve for parallax for foreground Galactic stars; (c) the weak-lensing survey goals
would benefit from the wealth of spectroscopic data provided by DESI and PFS in these fields.  The cons of this implementation would be: (a) the denser stellar regions near and including the disks of Andromeda and Triangulum would not be included (but could be done with a separate program); (b) the weak-lensing goals of HLWAS may be more difficult to achieve due to the closeness to the Galactic plane and the increased numbers of faint stars from the Andromeda and Triangulum halos; and (c) F062, which is not included in the HLWAS plan, would provide a better handle on photometric metallicities for Andromeda stars. 

\medskip

\noindent{\bf Option II:} A custom survey of Andromeda as part of the core plan provides the cleanest approach and could be optimally designed to accomplish multiple goals: (1) the halo survey described here; (2) a deep color-based survey of the Andromeda and Triangulum disks; (3) time domain surveys with different cadences in the disk and halo to identify variable stars. The total area is small, and the basic survey described here can be accomplished in $\sim$ 1 month. Adding these other components over a small area will only increase this commitment by a week of total time at the very most (the disk can be covered in $< 5$ Roman fields).  The main constraint is that the first epochs of the survey must be completed in the first year of Roman science operations, to ensure the longest possible time baseline for proper motion measurements. 

A third option may be to combine these; i.e., do a portion of the outer halo of Andromeda in the HLWAS and target the remainder of the footprint with a custom survey. 

\section{Trade Space and Expansion Options}

The current proposed survey baseline is for covering 500 sq. deg. in 2 filters during the first epoch and one filter in the second epoch. The {\underline{\it minimal}} requirement is to deliver 2 filters (for color information) and 2 epochs (for proper motion information). While it is possible that ground-based imaging (e.g., from Subaru/Hyper-SuprimeCam) could provide useful deep optical imaging, these data would suffer from confusion in the very central parts of the halo and in the disk of the galaxy, and would require the Roman data for source disambiguation (i.e., star/galaxy separation) and deblending. 

As a result, the only way to scale back the Roman survey proposed here without greatly impacting the science yield is to reduce the area. The 500 sq. deg. proposed area is defined to sample the Andromeda halo out to 200~kpc in certain directions, even beyond the imaging area covered by the PAndAS survey. The first scale-back option is to reduce the footprint to the $\approx$~350 sq. deg. imaged in the PAndAS survey (i.e., the jagged black outline in Fig.~\ref{fig:pandasmap}), which covers the Andromeda halo out to slightly less than $\sim150$~kpc and includes the region around Triangulum. The second scale-back option (not preferred) is to excise Triangulum from the survey. 

If the Roman mission lifetime is extended beyond the baseline, this provides an invaluable opportunity to add a third epoch. This would increase the time baseline for the survey and improve the accuracy of the proper motion measurements; if Roman has a lifetime similar to that of the Hubble Space Telescope, it would potentially increase the proper motion measurement accuracy to $\approx 3\mu$as/year (since $\sigma(PM)\propto t^{-1.5}$), or $\approx 11$~km/s at the distance of Andromeda. However, to maximize the future science return, it is {\underline{\it essential}} that the first epoch be obtained as early as possible during the Roman mission.

\bibliography{bibliography}{}
\bibliographystyle{aasjournal}

\end{document}

%% file: subauth.tex
\author[0000-0002-4928-4003]{Arjun Dey}
  \affiliation{NOIRLab, 950 N Cherry Ave, Tucson, AZ 85719}
\author[0000-0002-5758-150X]{Joan Najita}
  \affiliation{NOIRLab, 950 N Cherry Ave, Tucson, AZ 85719}
\author[0000-0001-5522-5029]{Carrie Filion}
  \affiliation{Johns Hopkins University}
\author[0000-0002-6800-5778]{Jiwon Jesse Han}
  \affiliation{Harvard-Smithsonian Center for Astrophysics, 60 Garden St, Cambridge, MA 02138}
\author[0000-0003-0256-5446]{Sarah Pearson}\thanks{Hubble Fellow}
\affiliation{Center for Cosmology and Particle Physics, Department of Physics, New York University, 726 Broadway, New York, NY 10003, USA}
\author[0000-0002-4013-1799]{Rosemary Wyse}
  \affiliation{Johns Hopkins University, 3400 N. Charles St, Baltimore, MD 21218, USA}
\author[0000-0001-7928-1973]{Adrien C. R. Thob}
  \affiliation{Department of Physics and Astronomy, University of Pennsylvania, 209 South 33rd Street, Philadelphia, PA 19104 USA}

%% file: auth.tex
%\author[0000-0002-4928-4003]{Arjun Dey}
%  \affiliation{NOIRLab, 950 N Cherry Ave, Tucson, AZ 85719}
%\author[0000-0002-5758-150X]{Joan Najita}
%  \affiliation{NOIRLab, 950 N Cherry Ave, Tucson, AZ 85719}
%\author[0000-0001-5522-5029]{Carrie Filion}
%  \affiliation{Johns Hopkins University}
%\author[0000-0002-6800-5778]{Jiwon Jesse Han}
%  \affiliation{Harvard-Smithsonian Center for Astrophysics, 60 Garden St, Cambridge, MA 02138}
%\author[0000-0003-0256-5446]{Sarah Pearson}\thanks{Hubble Fellow}
%\affiliation{Center for Cosmology and Particle Physics, Department of Physics, New York University, 726 Broadway, New York, NY 10003, USA}
\author[0000-0001-5261-4336]{Borja Anguiano}
  \affiliation{Department of Physics \& Astronomy, University of Notre Dame, Notre Dame, IN 46556, USA}
\author[0009-0001-2827-1705]{Miranda Apfel}
\affiliation{University of California, Santa Cruz}
\author[0000-0001-7214-3009]{Magda Arnaboldi}
  \affiliation{ESO, K. Schwarzschild str. 2, 85478 Garching, Germany}
  \author[0000-0002-5564-9873]{Eric F. Bell}
  \affiliation{Department of Astronomy, University of Michigan, 1085 S.\ University Ave., Ann Arbor, MI, 48109}
  \author[0000-0002-0740-1507]{Leandro {Beraldo e Silva}}
  \affiliation{Department of Astronomy, University of Michigan, 1085 S. University Ave, Ann Arbor, MI, 48109, USA}
\author[0000-0003-0715-2173]{Gurtina Besla}
  \affiliation{Steward Observatory, University of Arizona, 933 North Cherry Avenue, Tucson, AZ 85721, USA}
\author[0000-0002-9948-8442]{Aparajito Bhattacharya}
\affiliation{St. Xavier's College (Autonomous), 30 Mother Teresa Sarani Kolkata 700016 West Bengal, India}
\author{Souradeep Bhattacharya}
\affiliation{Inter University Center for Astronomy and Astrophysics, Ganeshkhind, Post Bag 4, Pune - 411007, India}
\author[0000-0002-0572-8012]{Vedant Chandra}
\affiliation{Center for Astrophysics $|$ Harvard \& Smithsonian, 60 Garden St, Cambridge, MA 02138, USA}
\author[0000-0003-1680-1884]{Yumi Choi}
\affiliation{NOIRLab, 950 N. Cherry Ave, Tucson, AZ 85719, USA}
\author[0000-0002-1693-3265]{Michelle L. M. Collins}
 \affiliation{Physics Department, University of Surrey, Guildford, GU2 7XH, UK}
\author[0000-0002-6993-0826]{Emily C. Cunningham}
\affiliation{Department of Astronomy, Columbia University, 550 West 120th Street, New York, NY, 10027, USA}
\author[0000-0002-1264-2006]{Julianne J. Dalcanton}
\affiliation{Department of Astronomy, University of Washington, Box 351580, U.W., Seattle, WA 98195-1580, USA}
\affiliation{Center for Computational Astrophysics, Flatiron Institute, 162 Fifth Ave, New York, NY 10010, USA}
\author[0000-0002-9933-9551]{Ivanna Escala}
\altaffiliation{Carnegie-Princeton Fellow}
\affiliation{Department of Astrophysical Sciences, Princeton University, 4 Ivy Lane, Princeton, NJ, 08544 USA}
\affiliation{The Observatories of the Carnegie Institution for Science, 813 Santa Barbara St, Pasadena, CA 91101}
\author[0000-0003-1183-701X]{Hayden R. Foote} \affil{Steward Observatory, The University of Arizona, 933 North Cherry Avenue, Tucson, AZ 85721, USA.}
\author[0000-0001-7934-1278]{Annette M. N. Ferguson}
\affiliation{Institute for Asronomy, University of Edinburgh, Blackford Hill, Edinburgh, UK EH9 3HJ}
\author[0000-0001-8203-6004]{Benjamin J. Gibson}
  \affiliation{Department of Physics and Astronomy, University of Utah, Salt Lake City, UT 84112, USA}
\author[0000-0001-9852-9954]{Oleg Y. Gnedin}
  \affiliation{Department of Astronomy, University of Michigan, 1085 S. University Ave, Ann Arbor, MI, 48109, USA}
\author[0000-0001-8867-4234]{Puragra Guhathakurta}
\affiliation{Department of Astronomy and Astrophysics, University of California Santa Cruz, 1156 High Street, Santa Cruz, CA 95064, USA}
\author[0000-0002-1423-2174]{Keith Hawkins}
 \affiliation{Department of Astronomy, The University of Texas at Austin, 2515 Speedway Boulevard, Austin, TX 78712, USA}
\author[0000-0003-1856-2151]{Danny Horta}
  \affiliation{Center for Computational Astrophysics, Flatiron Institute, 162 Fifth Ave, New York, NY 10010, USA}
\author[0000-0002-3292-9709]{Rodrigo Ibata}
  \affiliation{Universit\'e de Strasbourg, CNRS, Observatoire astronomique de Strasbourg, UMR 7550, F-67000 Strasbourg, France}
  \author[0000-0002-3204-1742]{Nitya Kallivayalil}
  \affiliation{Department of Astronomy, University of Virginia, 530 McCormick Rd., Charlottesville, VA 22904}
\author[0000-0001-9605-780X]{Eric W. Koch}
\affiliation{Center for Astrophysics $\mid$ Harvard & Smithsonian, 60 Garden St., 02138 Cambridge, MA, USA}
\author[0000-0003-2644-135X]{Sergey Koposov}
  \affiliation{Institute for Astronomy, University of Edinburgh, Royal Observatory, Blackford Hill, Edinburgh EH9 3HJ, UK}\affiliation{Institute of Astronomy, University of Cambridge, Madingley Road, Cambridge CB3 0HA, UK}
\author[0000-0003-3081-9319]{Geraint F. Lewis}
  \affiliation{Sydney Institute for Astronomy, School of Physics, The University of Sydney, NSW 2006, Australia}
\author[0000-0002-1775-4859]{Lucas Macri}
  \affiliation{NOIRLab, 950 N. Cherry Ave, Tucson, AZ 85719}
\author[0000-0001-7494-5910]{Kevin A. McKinnon}
  \affiliation{Department of Astronomy and Astrophysics, University of California Santa Cruz, 1156 High Street, Santa Cruz, CA 95064, USA}
\author[0000-0002-1793-3689]{David L. Nidever}
\affiliation{Department of Physics, Montana State University, P.O. Box 173840, Bozeman, MT 59717-3840}
\author[0000-0002-7134-8296]{Knut A.G. Olsen}
	 \affiliation{NOIRLab, 950 N. Cherry Ave, Tucson, AZ 85719}
\author[0000-0002-9820-1219]{Ekta~Patel}
\affiliation{Department of Astronomy, University of California, Berkeley, 501 Campbell Hall, Berkeley, CA, 94720, USA}
\author[0000-0003-1517-3935]{Michael S. Petersen}
\affiliation{Institute for Astronomy, University of Edinburgh, Royal Observatory, Blackford Hill, Edinburgh EH9 3HJ, UK}
\author[0000-0003-4030-3455]{Andreea Petric}
  \affiliation{Space Telescope Science Institute, 3700 San Martin Drive, Baltimore, MD 21210, USA}

\author[0000-0003-0872-7098]{Adrian~M.~Price-Whelan}
  \affiliation{Center for Computational Astrophysics, Flatiron Institute, 162 Fifth Ave, New York, NY 10010, USA}
\author[0000-0003-0427-8387]{R. Michael Rich}
  \affiliation{Department of Physics and Astronomy, UCLA, Los Angeles, CA, 90095-1547 USA}
\author[0000-0001-5805-5766]{Alexander H.~Riley}
\affiliation{Institute for Computational Cosmology, Department of Physics, Durham University, Durham DH1 3LE, UK}
\author[0000-0002-6839-4881]{Abhijit Saha}
\affiliation{NSF's National Optical Infrared Astronomy Research Laboratory, 950 North Cherry Avenue, Tucson, AZ 85719}
\author[0000-0003-3939-3297]{Robyn E. Sanderson}
  \affiliation{Department of Physics and Astronomy, University of Pennsylvania, 209 South 33rd Street, Philadelphia, PA 19104 USA}
\author[0000-0002-0920-809X]{Sanjib Sharma}
\affiliation{Space Telescope Science Institute, Baltimore, USA}

\author[0000-0001-8368-0221]{Sangmo Tony Sohn}
\affiliation{Space Telescope Science Institute, 3700 San Martin Drive, Baltimore, MD 21218, USA}
\author[0000-0001-6360-992X]{Monika~D.~Soraisam}
\affiliation{Gemini Observatory/NSF’s NOIRLab, 670 N.~A’ohoku Place, Hilo, HI 96720, USA}
\author[0000-0001-6516-7459]{Matthias Steinmetz}
  \affiliation{Leibniz-Institut f{\"u}r Astrophysik Potsdam (AIP), An der Sternwarte 16, 14482 Potsdam, Germany}
\author[0000-0002-6257-2341]{Monica Valluri}
\affiliation{University of Michigan, Department of Astronomy,1085 S. University Ave. Ann Arbor, MI 48109}
\author[0000-0003-4341-6172]{A.~Katherina~Vivas}
\affiliation{Cerro Tololo Inter-American Observatory/NSF’s NOIRLab, Casilla 603, La Serena, Chile}
\author[0000-0002-7502-0597]{Benjamin F. Williams}
\affiliation{Department of Astronomy, University of Washington, Box 351580, U.W., Seattle, WA 98195-1580, USA}
\author[0000-0002-3233-3032]{J. Leigh Wojno}
\affiliation{Max-Planck Institute for Astronomy, Königstuhl 17, D-69117 Heidelberg, Germany}

%\author[0000-0003-2229-011X]{Risa H.~Wechsler}
%  \affiliation{Kavli Institute for Particle Astrophysics \& Cosmology, P. O. Box 2450, Stanford University, Stanford, CA 94305, USA }
%  \affiliation{Department of Physics, Stanford University, 382 Via Pueblo Mall, Stanford, CA 94305, USA}
%\affiliation{SLAC National Accelerator Laboratory, Menlo Park, CA 94025, USA }
%\author[0000-0002-3569-7421]{Edward F.~Schlafly}
%  \affiliation{Space Telescope Science Institute, 3700 San Martin Dr, Baltimore, MD 21218, USA}
%\author[0000-0001-5999-7923]{Anand Raichoor}
%  \affiliation{Lawrence Berkeley National Lab, 1 Cyclotron Rd, Berkeley CA 94720}
%\author[0000-0003-2644-135X]{Sergey Koposov}
%  \affiliation{EDI CAM}
%\author[0000-0003-0715-2173]{Gurtina Besla}
%  \affiliation{University of Arizona}
%\author[0000-0002-5042-5088]{David J Schlegel}
%  \affiliation{Lawrence Berkeley National Lab, 1 Cyclotron Rd, Berkeley CA 94720}
%\author[0000-0002-5564-9873]{Eric F.~Bell}
%  \affiliation{Department of Astronomy, University of Michigan, 1085 S. University Ave, Ann Arbor, MI, 48109, USA}
%\author[0000-0002-6257-2341]{Monica Valluri}
%  \affiliation{Department of Astronomy, University of Michigan, 1085 S. University Ave, Ann Arbor, MI, 48109, USA}
%\author[0000-0001-8274-158X]{Andrew Cooper}
%  \affiliation{National Tsing Hua University}

% Unique acks:
\newcommand{\allacks}{
Acknowledgements
}